\documentclass[fleqn,10pt]{wlscirep}
\usepackage[utf8]{inputenc}
\usepackage[T1]{fontenc}
\usepackage{lineno}
\usepackage{booktabs}
\usepackage{todonotes,soul,comment}

\title{GameVibe: a multimodal affective game corpus}

\author[1,$\dag$]{Matthew Barthet}
\author[1,$\dag$]{Maria Kaselimi}
\author[1,$\dag$]{Kosmas Pinitas}
\author[1,$\dag$]{Konstantinos Makantasis}
\author[1,$\dag$]{Antonios Liapis}
\author[1,$\dag$]{Georgios N. Yannakakis}

\affil[1]{Institute of Digital Games, University of Malta, Malta}

\affil[*]{corresponding author(s):Author1N Author1S (author1n.author1s@um.edu.mt)}

\affil[$\dag$]{these authors contributed equally to this work}

\begin{abstract}
As online video and streaming platforms continue to grow, affective computing research has undergone a shift towards more complex studies involving multiple modalities. However, there is still a lack of readily available datasets with high-quality audiovisual stimuli. In this paper, we present GameVibe, a novel affect corpus which consists of multimodal audiovisual stimuli, including in-game behavioural observations and third-person affect traces for viewer engagement. The corpus consists of videos from a diverse set of publicly available gameplay sessions across 30 games, with particular attention to ensure high-quality stimuli with good audiovisual and gameplay diversity. Furthermore, we present an analysis on the reliability of the annotators in terms of inter-annotator agreement.
\end{abstract}

\begin{document}

\flushbottom
\maketitle
\thispagestyle{empty}

\section*{Background \& Summary}
\label{sec:background}

Affective Computing (AC) is the interdisciplinary field that refers to the study of human emotions and the development of tools and technologies that learn, interpret and perceive these emotions \cite{picard2001toward}. The availability of large-scale corpora comprising affect manifestations elicited through appropriate stimuli is critical for AC. However, manifestations of affect are often highly subjective and difficult to assess; each individual's interpretation of a stimulus is influenced by factors such as preferences, memory, biases, systematic errors, and expectations \cite{sciutti2022affective,yannakakis2018ordinal}. Furthermore, emotions are a dynamic phenomenon and participants' reactions to similar stimuli may shift over time. Therefore, as AC advances and deep learning methods require an increasing amount of data and scale, it becomes important to design and implement experimental protocols that maximise the reliability of collected labels of large-scale affect corpora. 

Video games are a hugely popular type of digital media, offering a unique form of human-computer-interaction (HCI) due to their rich content and interactive nature \cite{yannakakis2023affective}. The global video game industry has seen substantial expansion and is already among the fastest growing subsectors within the broader entertainment sector \cite{goh2023unravelling}. Thus, using games as a test-bed to reveal the intricacies of the HCI loop is one of the most promising ways to analyse human behaviour and experience, at scale. Studying the behaviour of players and the experience of play can, in turn, be used to improve the technical aspects of game development, and to contribute to more engaging and personalised game experiences \cite{lopes2017ranktrace, kotsia2013affective}. 

The literature already contains a growing number of diverse affect corpora that vary in terms of modalities of user input considered, annotation methods, affect stimuli, and number of participants among other factors. Some notable examples of modalities recorded include audiovisual data \cite{maman2020game, doyran2021mumbai}, physiological signals \cite{kollias2022abaw}, and facial expression data \cite{mollahosseini2017affectnet}. \emph{First-person} annotation protocols ask the annotator to label their own experience, and have been employed in several affect corpora, including MazeBall \cite{yannakakis2010towards}, PED \cite{karpouzis2015platformer}, FUNii \cite{beaudoin2019funii}, and MUMBAI \cite{doyran2021mumbai}. When employing a \emph{third-person} protocol, instead, we ask annotators to label the experience of another person, such as in the RECOLA  \cite{ringeval2013introducing}, LIRIS-ACCEDE \cite{baveye2015liris}, AFF-Wild \cite{kollias2022abaw}, AffectNet \cite{mollahosseini2017affectnet}, and SEWA \cite{kossaifi2019sewa} corpora. The provided labels can be \emph{discrete} (e.g. categories, scales) such as in GAME-ON \cite{maman2020game} and BIRAFFE2 \cite{kutt2022biraffe2} or \emph{continuous} traces as in RAGA \cite{granato2020empirical} and MUMBAI \cite{doyran2021mumbai}. 

Contemporary affect corpora are gradually deviating from the controlled setting of in-lab experiments \cite{ringeval2013introducing,forgas1984influence,pinilla2020induced} to real-life (or \emph{in-the-wild}) scenarios \cite{kutt2022biraffe2, park2020k, mollahosseini2017affectnet} in an attempt to elicit more natural user behaviours and manifestations of affect. Hiring annotators from crowdsourcing platforms has also been gaining interest in an attempt for AC to scale. Untrained crowd workers in uncontrolled settings, however, can be unreliable, resulting in corpora of questionable, or even limited, value \cite{barthet2023knowing}. It is fair to say that all affect corpora are characterised by subjectivity \cite{martinez2023reliability, miranda2018amigos, kollias2019deep}, since consensus among annotators is not necessarily required. Whilst some studies attempt to mitigate this issue of variability (e.g. by using mood induction \cite{lench2011discrete}) for better performance in downstream tasks, doing so may hamper the generalizability and reliability of results when deployed in uncontrolled real-life scenarios. Motivated by these issues, we seek to provide an affect corpus with validated annotator reliability and sufficient diversity in the stimuli to promote generalizability in downstream tasks. We aim to accomplish this by following a quality assurance protocol \cite{barthet2023knowing}, and collect stimuli which present rich contextual variety within a single domain to promote deeper research into the generalizability of affect predictions.

Regarding games as affect stimuli, the players' (or viewers') perceptions of a gameplay video as stimulus are inextricably linked to the game genre, the form of interfacing, the game's objective, the number of players, and potential social aspects  \cite{yannakakis2023affective}. We specifically focus on the First-Person Shooter (FPS) genre of games in this paper, for many reasons. FPS games have been popular among players since the 1990s \cite{morris2003wads}. Moreover, FPS games usually have high-quality graphics and audio, and a rich variability of stimuli due to the vast number of FPS games developed over the years. FPS games tend to be highly stimulating for both players and viewers, and share many fundamental gameplay elements whilst simultaneously employing very different styles (in terms of both art and gameplay). In addition, the FPS genre has been hugely popular on live-streaming platforms such as Twitch for several years. Indicatively, during December 2023, 4 of the top 10 live-streamed categories were FPS games, with over 220 million viewer hours watched in just 30 days (\url{https://www.twitchmetrics.net/games/viewership}, accessed on 16/12/2023). We argue that collectively these factors make FPS games the ideal genre for studying HCI, especially for understanding viewer engagement. This makes research in FPS affect modelling highly useful for content creators, streamers, and game designers. 

With this in mind, we introduce \emph{GameVibe}, a novel multimodal affect corpus of viewer engagement for FPS game videos. This corpus consists of 2 hours of high-quality audio and visual data from 30 different FPS games extracted and curated from publicly available ``Let's Play'' videos on YouTube. Data collection involved 20 annotators, consisting of trained researchers, as well as postgraduate and undergraduate students. Affect traces were provided in the form of unbounded, time continuous signals using the RankTrace annotation tool \cite{lopes2017ranktrace} on the PAGAN platform \cite{melhart2019pagan}. As part of the corpus, we provide the raw videos used as stimuli during data collection, as well as latents extracted using pretrained foundation models for visuals (VideoMAE \cite{tong2022videomae} and MVD \cite{wang2023masked}),  and audio (BEATS \cite{chen2022beats}) for use in downstream tasks. Furthermore, we include quality assurance data on each annotator to give a better understanding of the reliability and validity of the annotations in the dataset. Finally, we test the reliability and validity of the dataset in terms of inter-annotator agreement. 

Our in-lab data collection protocol and annotator quality assurance tools ensure that the GameVibe corpus consists of annotations which reliably approximate the ground truth. This corpus can be applied to a variety of downstream tasks, such as modelling engagement using the provided latent representations extracted from pre-trained models for vision and audio. Initial studies using GameVibe have shown that models can reliably predict viewer engagement in unseen clips when trained on clips of the same game \cite{pinitasvarying}, which can be useful for game developers to assess the quality of new content. Another downstream task which benefits multimodal affect modelling research tackles the generalizability of affect models \cite{camilleri2017towards} through the diversity in stimuli of this corpus. Our generalizability study on building engagement models which predict labels in unseen games using few-shot learning has shown promising results \cite{pinitas2024across}, and marks further progress towards the ultimate goal of general models of affect. The release of this corpus is expected to contribute both in engagement modelling and multimodal interaction research more broadly than these initial studies.

    

\section*{Methods}\label{sec:methods}
Measuring the experience of digital game enjoyment and accurately identifying which game elements engage players are important goals for both the study of user experience in games and the development of better games \cite{yannakakis2023affective}. In this section, we introduce the FPS affect corpus GameVibe, which was solicited to provide a rich, multimodal dataset of elicitors for viewer engagement. The collected dataset includes: (a) synchronised frames and audio per game, (b) extracted latent representations from audiovisual data, (c) engagement annotation traces per participant (in raw \& processed form), (d) participants' replies to demographic surveys (anonymised data). The novelty of the dataset lies in the fact that engagement annotations are user-specific; by exploiting the different opinions and perspectives of different users for the same game, GameVibe enables the training of fair affect models that are robust to new users and new games.

\begin{figure}[t]
\centering
\includegraphics[width=\textwidth]{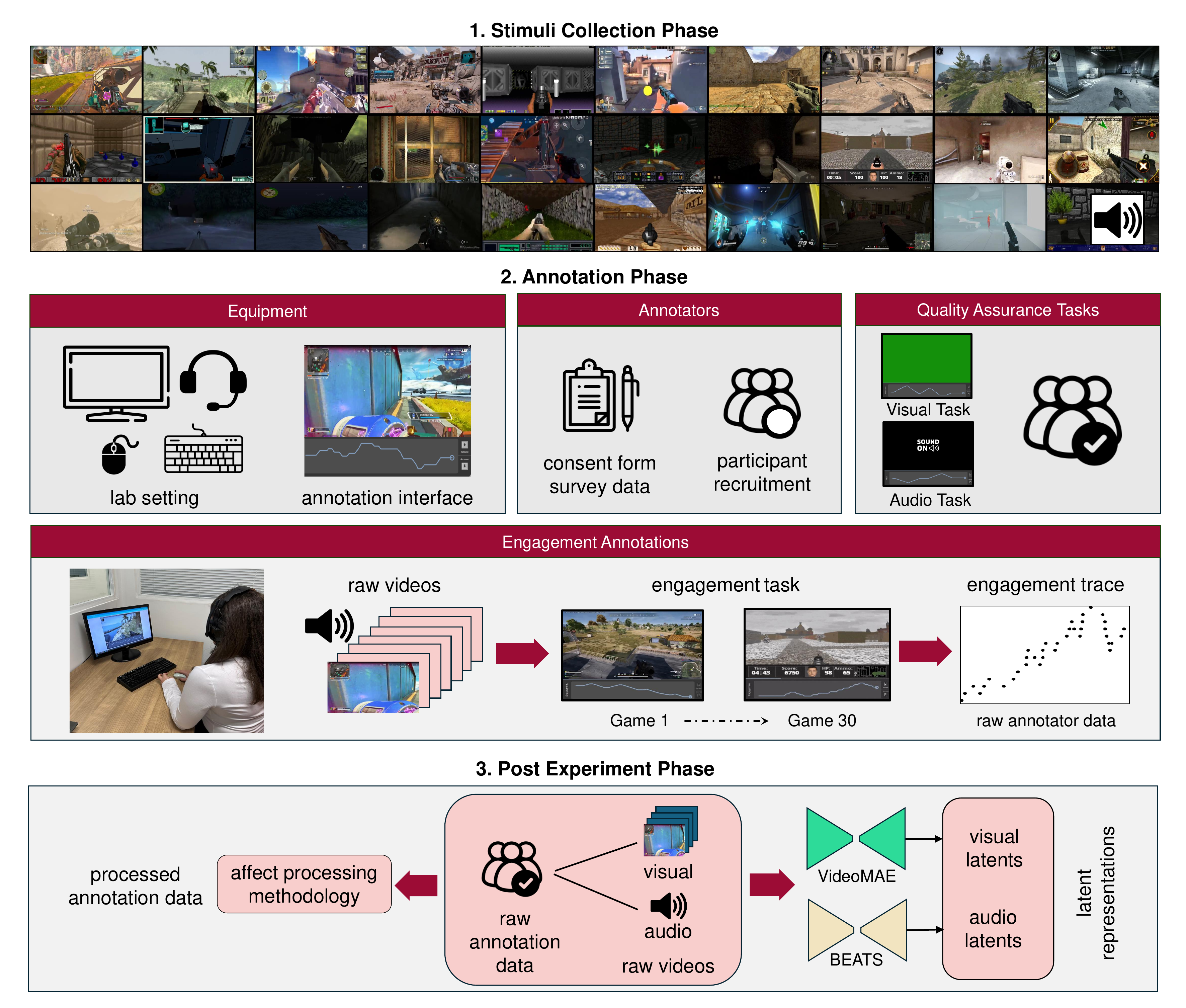}
\caption{High-level overview of the experimental protocol used for data collection in this study.}
\label{fig:methodology}
\end{figure}

In this section, we provide a detailed description of the process we followed to build GameVibe. The core phases, as illustrated in Fig. \ref{fig:methodology}, are as follows:

\begin{itemize}
  \item \textbf{Design phase:} we outline the objectives, tools, methodologies, and parameters of the study. 
  \item \textbf{Stimuli collection phase:} we choose and curate video content from diverse FPS games.
  \item \textbf{Annotation phase:} before collecting annotations of viewers' emotional states, we recruit participants, procure and prepare the equipment, and develop informed consent forms to uphold ethical standards. In addition, we conduct Quality Assurance tests to measure participants' reliability. We detail all the above below.
  \item \textbf{Post-experiment phase:} we assess the quality of the dataset, process and analyse the collected data and document our findings. This ensures the dataset's efficacy and utility for subsequent research endeavours.
\end{itemize}

\subsection*{Design phase}
This phase involves defining the scope and purpose of the dataset, determining the specific types of data to be collected, establishing data sources and acquisition methods, and devising a structure for organizing and storing the data. Additionally, factors such as data format, granularity, quality, and potential biases were considered. Beyond research standards, collaboration with domain experts from the game industry is essential to understand requirements and ensure that the dataset aligns with intended use cases. In this work, we draw from our experiences during our long-term collaboration with \emph{Ubisoft's Massive Entertainment}, with whom we co-designed an effective experimental protocol for collecting annotations and training affect models in \emph{Tom Clancy's The Division 2} \cite{pinitas2023predicting}. Following similar practices, we ensure that our dataset closely aligns with both research and industry goals. Furthermore, privacy and ethical considerations were carefully weighed to safeguard sensitive information and uphold ethical standards. In this study, we exploit already tested and reliable tools---based on self-reporting---to measure a game's engagement level as perceived by different annotators. We also rely on well-established methodologies for gathering emotion-related datasets, specifically the PAGAN data collection framework \cite{melhart2019pagan} and the RankTrace annotation tool \cite{lopes2017ranktrace}. Furthermore, our core objective was to solicit reliable engagement annotations across a wide variety of contexts to allow for further research into the generalizability of affect models across multiple annotators and stimuli.

\begin{figure}[t]
\centering
\includegraphics[width=\textwidth]{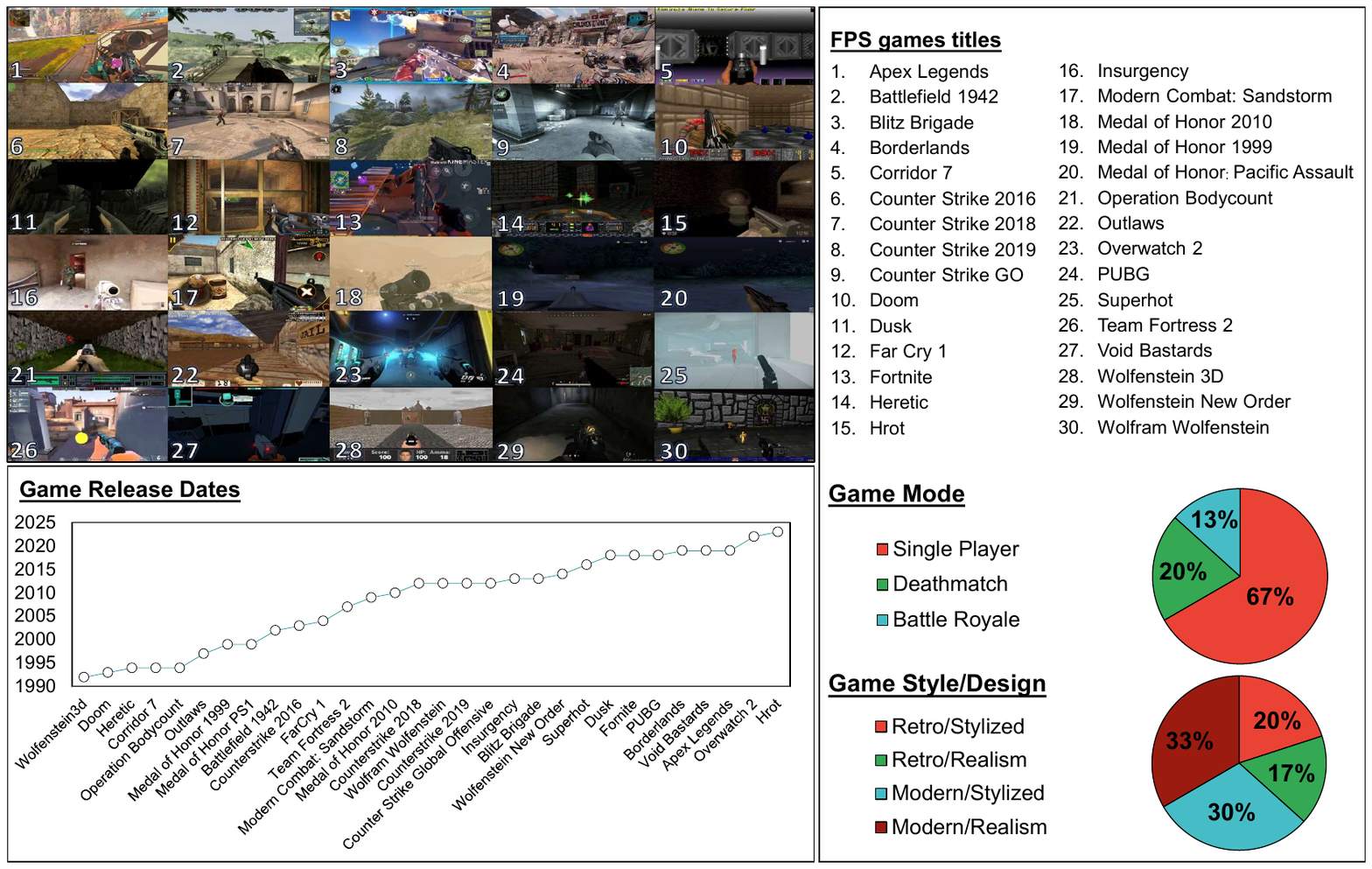}
\caption{Details of the GameVibe audiovisual stimuli. The figure illustrates screenshots from the 30 different FPS games that are annotated for engagement (top left), the release date of each title in ascending order (bottom left),  the names of each FPS title (top right) and the proportion of games in terms of game modes, game styles and game designs (bottom right).}

\label{fig:stimuli}
\end{figure}

\subsection*{Stimuli collection phase}
As discussed in \textbf{Background \& Summary}, we chose 30 FPS games as affect stimuli in order to encompass the widest possible range of audiovisual styles and design aspects (including modes of gameplay and winning conditions). In terms of \emph{game style,} we distinguish between games featuring a ``realistic'' vs. ``stylized'' art style. In terms of \emph{game era,} we pick games of both ``modern'' and ``retro'' styles (Fig. \ref{fig:stimuli}). For example, a game such as \textit{Overwatch 2} (game 23 in Fig. \ref{fig:stimuli}) falls under the modern, stylised art style, whereas \textit{Wolfenstein} (game 28 in Fig. \ref{fig:stimuli}) would fall under the retro gameplay with a realistic art style. For distinguishing across \emph{game design} patterns, we made an effort to select games with different game modes or ``sub-genres'' within the FPS space, such as ``Battle Royale'' (e.g. \textit{PUBG}), ``death match'' (e.g. \textit{Counter Strike}), and ``single-player'' (e.g. \textit{Doom}) games. The distribution of these design patterns is visualised in Fig. \ref{fig:stimuli}, where we can see single player games make up 67\% of the stimuli selected, with death match and Battle Royale making up 20\% and 13\% respectively. We hope that the richness in variety and multimodality of the dataset will empower affect models to generalise better to unseen games when predicting viewer engagement, and help us understand how game design and graphical style impact human responses to these stimuli. 

Within each game, when picking the videos to be included in the dataset, we ensured that the content consisted of primarily gameplay, meaning that there are no videos with more than 15 seconds of non-gameplay footage (e.g. menu screens, cut scenes or transition animations). Furthermore, we made sure to only use videos with audio consisting solely of in-game sounds and that did not contain any player or streamer commentary. Based on these criteria, we acquired the videos for each of our chosen FPS games from ``Let's Play'' videos on YouTube, ensuring the videos were at least 5 minutes long in order to have enough content to extract multiple short videos from. Each video was then trimmed and broken down into four separate videos of one minute each, meaning each FPS game in the dataset has 4 distinct one-minute videos in the dataset.

In order to provide annotators with a variety of stimuli, each annotation \textit{session} contains 30 of the aforementioned 1-minute videos, one for each game (i.e. 30 games per session). The videos inserted into each session were selected by randomly sampling each game's set of videos without replacement. As a result, each session contains a unique set of 30 videos, which were then shown to annotators in a randomised order. This means the sessions took the form of a 30-minute sequential annotation task. Notably, the annotators were permitted to pause the task at any time or take breaks in between each video in order to minimise user fatigue. As an indicative example, \textit{Session 1} and \textit{Session 2} each contains thirty different gameplay stimuli (from the same pool of 30 games shown in Fig.~\ref{fig:methodology}) which are 1 minute each. Importantly, we consider videos from different sessions as independent stimuli despite originating from the same game, since the events depicted (and thus the perceived player engagement) are different---and differently timed---between gameplay videos from the same FPS game, or similar games.

\subsection*{Annotation phase}

\paragraph{Participant recruitment.}
We recruited 20 annotators among members of the University of Malta via convenience sampling. Annotators included research staff, B.Sc. students in fields relevant to digital games (including psychology and artificial intelligence), as well as M.Sc. students in Digital Games and PhD students in games research. Participants signed up for the study after being invited through the University's mailing list, and booked their slot at their own convenience. Recruitment and data collection started during March 2023 and was carried out throughout the year. Participants were also offered a €15 voucher as compensation upon completion of their participation in the study. 

\paragraph{Laboratory settings.}
All participants performed the annotation in the same room and light conditions at the Game Lab of the Institute of Digital Games, using the same machine and input/output devices (screen for visual stimuli, headphones for auditory stimuli, and a mouse with a scroll wheel for annotation). All participants were given a thorough introduction to the annotation task by at least one researcher involved in this work, who remained available during the entire annotation period for assistance and questions. Once the engagement annotation was completed (i.e. a task lasting approximately 30 minutes per participant), the participants were asked to complete a survey, thanked for their participation and exited. 

\paragraph{Research ethics.}
The above process was approved by the Institutional Review Board (IRB) of the University of Malta (\url{https://www.um.edu.mt/research/ethics/researchethicsatum/}, accessed September, 2024) before the annotation process commenced. Participants were required to sign a physical consent form prior to the experiment, agreeing to take part in the experiment and to have their annotation data and survey responses recorded and stored, as is common practice \cite{miranda2018amigos}. Participants were informed that they could halt the experiment at any time and leave without issue should they not wish to continue. While participants' demographic questions in the post-survey questionnaire (see below) included personal data, we preserved anonymity by using IDs rather than annotator names and ensured that such questions could not be used to identify the participants.

\paragraph{Survey data.} 
At the end of the annotation experiment, participants were required to fill in a short survey on their demographic information (age range, ethnicity, gender, handedness, education level), familiarity with video games, familiarity with FPS games, familiarity with video annotation as well as their favourite game. Most participants were between 25 and 35 years old (42\%), while 32\% were between 18 and 25 years and 21\% were between 35 and 45 years old; 1 participant was over 45 years old. Almost all participants (97.4\%)  were Caucasian, and the vast majority (89.5\%) were right-handed. Furthermore, 75\% of participants were male, 20\% were female and 5\% identified as non-binary. In terms of familiarity with games or video annotation, most participants were very familiar with games (average Likert score 4.25 out of 5) and slightly less so with FPS games (average Likert score 3.55 out of 5). Most participants were not very familiar with video annotation of affect data (average Likert score 2.65 out of 5), although some outliers existed (6 annotators rated their familiarity at 4 or 5 on a 5-point Likert scale). Finally, participants mentioned many different favourite games, with 7 out of 20 participants mentioning games which were first-person or third-person shooters.

\paragraph{Annotation protocol and tools.}
As mentioned, each participant was assigned to a session and asked to annotate their engagement for a randomised sequence of 30 videos. The random video order was imposed to minimise carry-over effects between stimuli, as has been established in the literature \cite{sharma2019dataset}. Annotation tasks were carried out using the PAGAN annotation platform \cite{melhart2019pagan} using the RankTrace annotation interface \cite{lopes2017ranktrace}. RankTrace allows for annotation in a continuous and unbounded fashion, and has been used extensively to collect reliable ground truths in an ordinal fashion \cite{yannakakis2018ordinal} (see equipment in Fig. \ref{fig:methodology}).

Before annotating gameplay videos, participants were tasked with annotating two Quality Assurance (QA) tasks, which were set up as independent PAGAN projects and linked together. Such QA tasks measure the ability to annotate a simple objective task where the ground truth signal is known. Such QA tasks have been proven to reliably predict annotator reliability in a previous study which used the first two sessions of this dataset \cite{barthet2023knowing} as well as other stimuli \cite{burmania2015increasing}. The first task is a visual task which requires the participant to annotate their perceived changes in brightness of a video showing a green screen, inspired by a previous study \cite{narayanan2020green}. The second task is an auditory task, which requires participants to annotate their perceived changes in pitch whilst listening to an oscillating sound wave. These two stimuli are included in the dataset along with their respective annotations to provide a measure of our annotators' reliability on simple objective tasks, and by extension a better insight into their reliability in our engagement study.

When participants were ready to annotate gameplay videos, they were provided the following definition of \textit{viewer engagement}: \emph{``A high level of engagement is associated with a feeling of tension, excitement, and readiness. A low level of engagement is associated with boredom, low interest, and disassociation with the game''}. We emphasised that participants annotated in a \textit{first-person} manner, i.e. they annotated their own engagement as a viewer and not their perceived engagement of the player. Based on this definition, participants were instructed to scroll up on the mouse wheel whenever they felt an increase in their engagement, down when they felt a decrease, and to remain idle (i.e. no scrolling) if their engagement remained unchanged. Since RankTrace annotations are unbounded, mouse scroll wheel offers an ideal---and intuitive---interface to collect such data.

\paragraph{Collected annotation data.}
As shown in Fig. \ref{fig:process}(b), the dataset consists of 4 sessions of annotated footage, producing a corpus of gameplay videos of a total duration of 2 hours. Each of the 4 sessions contains 30 different 1-minute videos (i.e. one video from each of the 30 game titles), which are presented to the annotator in random order. This means that for each session we collect 150 annotation traces, with every video being annotated by 5 participants. In total, this amounts to 600 engagement annotation traces for 120 gameplay videos (of 30 different game titles), annotated by 20 participants. The data are contained in the dataset repository \cite{game_vibe_osf}, detailed under \textbf{Data Records}.

\paragraph{Quality assurance and reproducibility of the GameVibe affect annotations.}
The dataset was created with high-quality videos from numerous FPS games, ordered randomly. The random order is commonly used to prevents bias from ordering effects \cite{d2018affective}, while the varied short video stimuli minimise user fatigue. The same room, equipment and light conditions were kept during the process for consistency of the annotation experience and data collection instruments (i.e. mouse wheel). Instructions were provided to the annotators to avoid gross errors during the annotation process. Finally, the dataset includes data from two QA tasks completed by annotators prior to the engagement study. Performance of annotators in such QA tests could be used in future studies to filter inconsistent annotators; however, in this paper, we include all annotators in the \textbf{Technical Validation} below.

\subsection*{Post-experiment phase}

\begin{figure}[t]
\centering
\includegraphics[width=\textwidth]{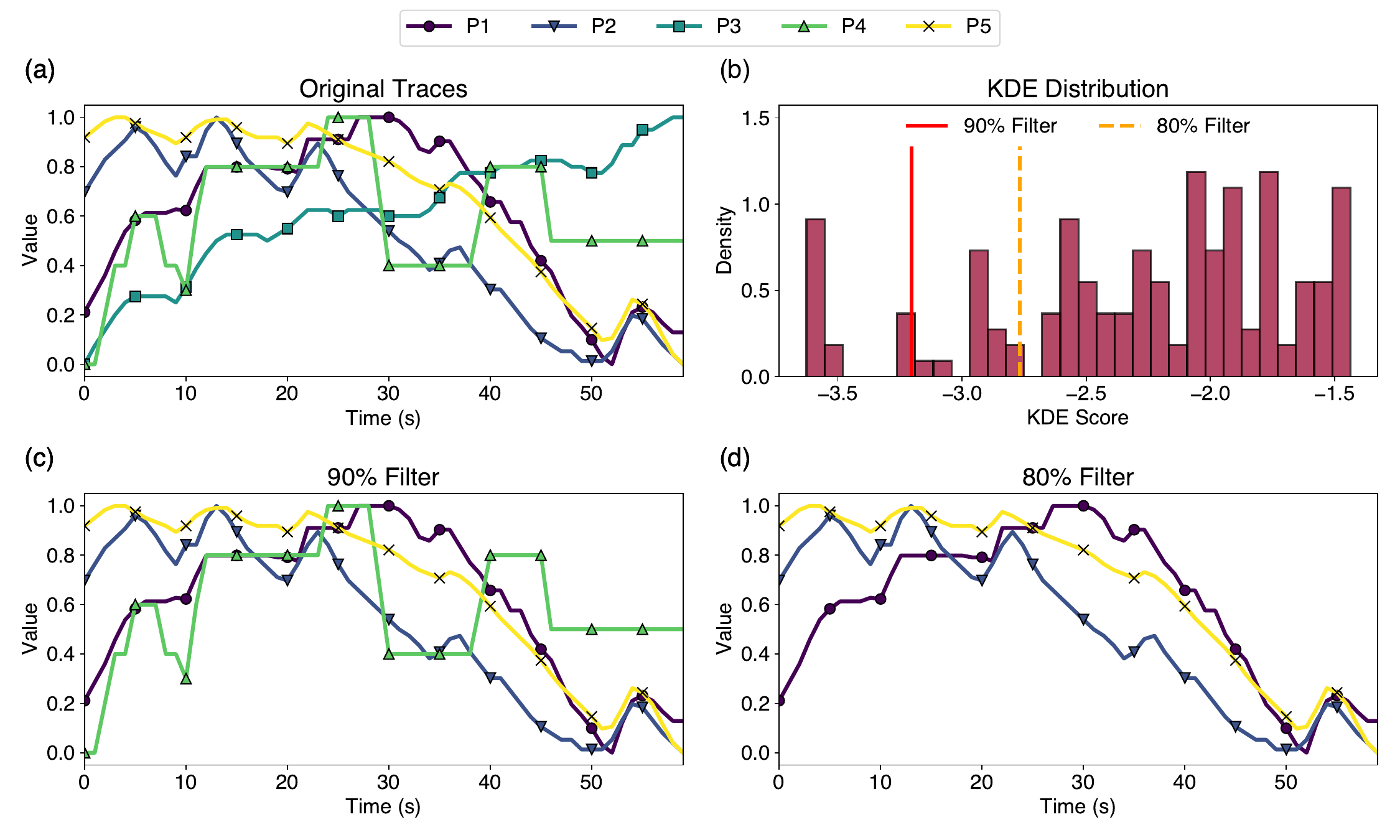}
\caption{An example of outlier detection using the game \emph{Wolfenstein 3D} (Apogee Software, 1992) from Session 1. Figure 3(a) depicts the unfiltered traces from five annotators. The frequency distribution can be seen in Figure 3(b), depicting the signals' KDE score used to create the outlier filters. The red and orange lines depict the 90\% filter and 80\% filter thresholds, respectively. Annotators to the right of both filters are considered inliers and are not removed. Figure 3(c) shows the first filter applied removing one outlier (P3), and Figure 3(d) shows the strictest filter applied removing two outliers (P3, P4). }
\label{fig:filtering}
\end{figure}

\paragraph{Audiovisual stimuli data processing.}
In this study, we exploit large-scale pretrained models to extract latent representations in visual data using (a) Video Masked Autoencoders (VMA) \cite{tong2022videomae} and (b) Masked Video Distillation (MVD) \cite{wang2023masked} models for visual representation learning. VMA is a model for robust representation learning for video data by utilising a masked autoencoder architecture \cite{tong2022videomae}. MVD provides efficient knowledge distillation by leveraging the principles of masked attention mechanisms to distil knowledge from a teacher network to a student network \cite{wang2023masked}. For audio data, we leverage the BEATS model proposed by Chen \textit{et al.} \cite{chen2022beats}, which serves as a self-supervised architecture designed for audio processing tasks. Given that FPS games have a high representation fidelity (regardless of art style) with the real world, we hypothesised that the above pretrained models can identify and classify different visual elements or sounds present within the audiovisual stimuli across different FPS games.

\begin{figure}[t]
\centering
\includegraphics[width=1\columnwidth]{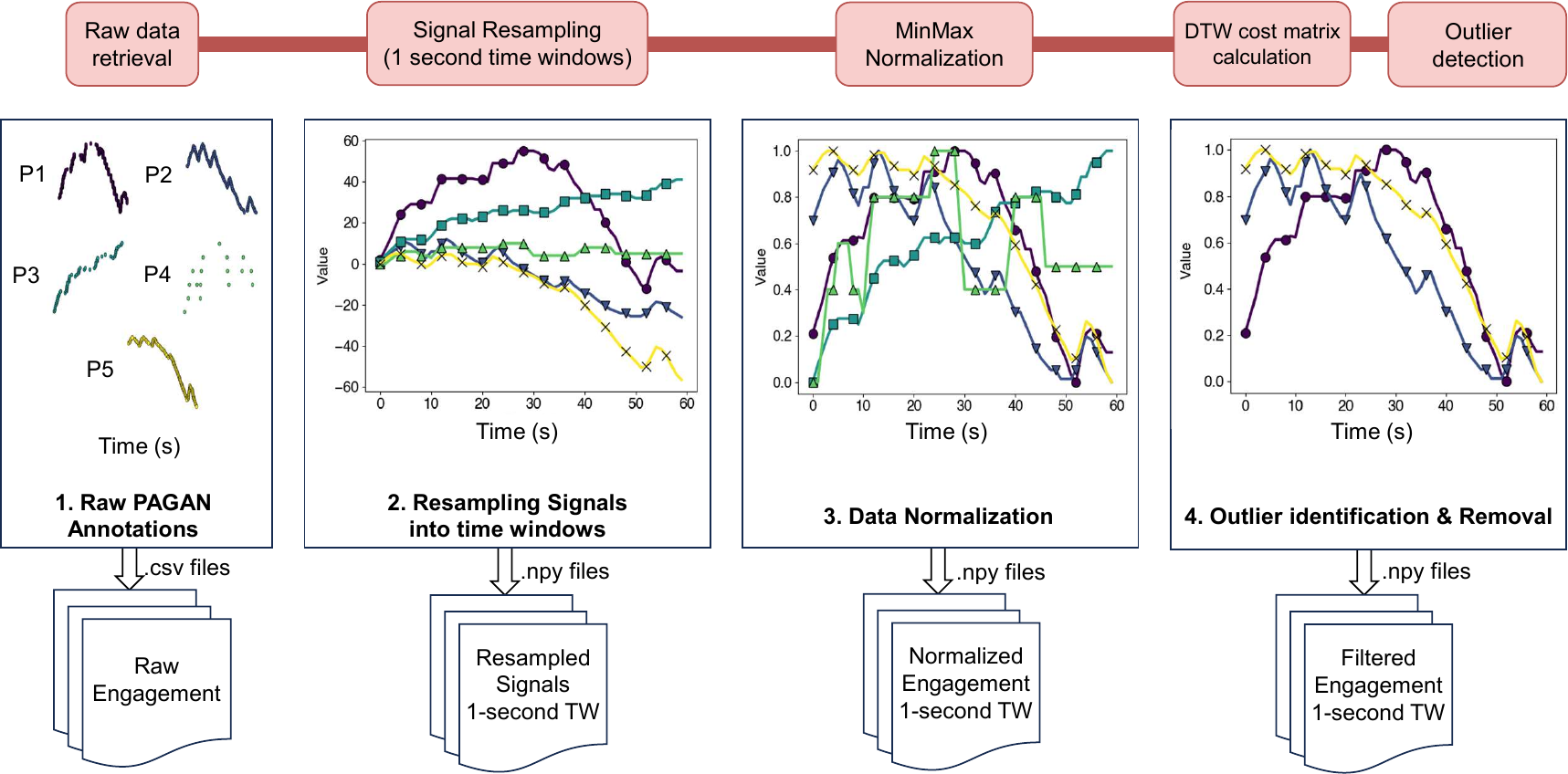}
\caption{GameVibe's post-processing pipeline, visualising the transformed annotations and the output files of each stage. }
\label{fig:process}
\end{figure}

\paragraph{Affect processing methodology.}
The raw discrete values that indicate changes in engagement are generated from each PAGAN project and require a number of pre-processing steps before they are suitable for use in downstream tasks. Raw data from PAGAN is a list of values paired with a timestamp for when that value was entered by the participant (after scrolling on the mouse wheel), stored in a CSV file for each session. This data is converted into a time continuous signal by interpolating the values into a series of 1-second time windows (see Step 1 in Fig. \ref{fig:process}). We also ensure that the annotator has made at least one input to the signal to be considered a valid signal; flat signals (with no changes) are discarded from the dataset, however no such instances were recorded. Finally, we combine the data into a single Python \emph{pickle} file, detailed under \textbf{Data Records}.

As RankTrace \cite{lopes2017ranktrace} was used to record engagement using a mouse scroll wheel, the raw data collected are unbounded (i.e., have no minimum or maximum values). In order to facilitate comparison across annotators, we include a normalised version of our dataset where engagement is normalised to [0,1] via Min-Max normalisation on a per-trace basis (see Step 2 in Fig. \ref{fig:process}). In our processing library, we also include the option to perform moving average smoothing on the signals in the dataset, or to change the size of the time window of interpolated signals. We selected a 1-second resampling rate, considering the trade-off between increasing data volume and detecting variations in engagement. Larger time windows or moving average filters would result in a smoother signal at the cost of signal accuracy and resolution. We offer the raw annotation traces from PAGAN in the dataset (see below) for further experiments with other processing methods. The balance of the collected data is highly dependent on said processing methods, and therefore we do not conduct an in-depth study on the engagement distributions for each stimulus. As an indication of the label balance, the corpus consists of 15.7\% increases, 13.4\% decreases and 70.9\% stable engagement levels when the raw data is processed into ordinal signals \cite{yannakakis2017ordinal} using 1 second time windows.

Since this study handles subjective annotations, where no concrete ground truth can ever be identified, two strategies are common to establish a consensus. One strategy is to embrace the potentially high variance in the dataset and use this information to identify regions of high vs low agreement \cite{DyNaMoS}, or even model it as part of the training process to improve performance \cite{rizos2019modelling}. Another strategy is to empirically analyse the data and use established techniques such as outlier detection \cite{boukerche2020outlier} to remove problematic annotators, or factor in annotator consensus \cite{parthasarathy2016using}. We follow the latter strategy, detecting and removing outliers in the dataset to improve the quality of the ground truth signal derived in any downstream tasks (see Step 3 in Fig. \ref{fig:process}). 
 
For filtering outliers in each video in a session (see Step 4 in Fig. \ref{fig:filtering}), we use Dynamic Time Warping \cite{DTW}  (DTW) to create a distance matrix of the normalised annotation signals from each participant. We chose DTW as a distance measure due to its proven use in similar studies \cite{AGAIN} and its ability to focus on comparing the shape of the signals (i.e. an ordinal distance measure) whilst also factoring out time shifts between participants due to different lags in their annotation \cite{mariooryad2014correcting}. For example, for the video of \textit{Wolfenstein 3D} in Session 1, we calculate the pairwise DTW distances between all possible pairs of annotators and assemble them into a 5$\times$5 DTW distance matrix. We use the minimum distance to the nearest participant to judge whether a participant is a singular outlier (and should therefore be removed), or part of a cluster of two or more annotators (and should therefore remain in the dataset). This process involves setting a threshold to determine what will be considered an in/outlier. Rather than using a static threshold which would require hand-picking for each session, we use the non-parametric Kernel Density Estimation (KDE) for outlier removal to create a threshold, which is visualised in Fig. \ref{fig:filtering}. This involves calculating all the pairwise KDE scores between signals for all videos in a session and sorting them in ascending order. We then test two filters, a strict filter which only includes signals with a nearest-neighbour distance within the best 80\% of KDE scores, and a more relaxed filter which includes signals with a nearest-neighbour distance within the best 90\% of KDE scores. Remaining signals are considered outliers and are removed: examples of removed outliers for \textit{Wolfenstein 3D} are shown in Fig. \ref{fig:filtering}.

\begin{figure*}[t]
\centering
\includegraphics[width=\textwidth]{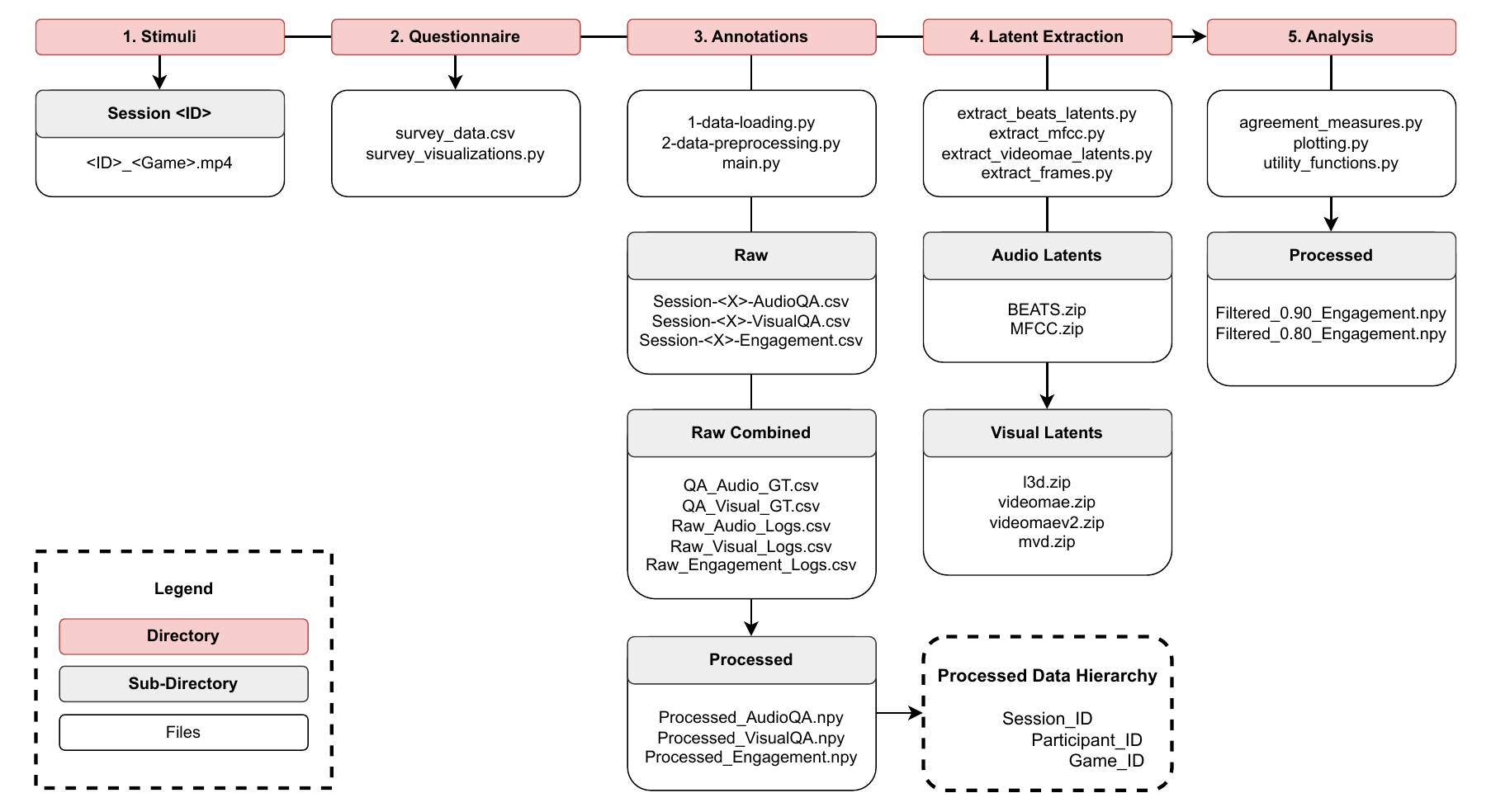}
\caption{Diagram overview of the dataset and its structure.}
\label{fig:datarecords}
\end{figure*}

\section*{Data Records}
The dataset and files described in this section are publicly available and can be found on our Open Science Framework repository \cite{game_vibe_osf}. We provide annotation data in CSV format and the respective video stimuli in \textit{.mp4} format: both formats can be processed in most software packages or programming language. We provide Python files for data processing and feature extraction. The \texttt{GameVibe\_README.md} file details the organisation of the dataset, explaining the structure, naming convention, and specific contents of each file.

The dataset is split into five directories as depicted in Fig. \ref{fig:datarecords}. The ``\textit{Stimuli}'' directory contains the 1-minute game videos organised into subdirectories by session and named according to the game in each video. The ``\textit{Questionnaire}'' directory contains participants' responses to the post-experiment questionnaire in CSV format, and a Python script to process the responses and create visualisations. The ``\textit{Annotations}'' directory contains the annotation data returned from PAGAN and the Python scripts required to process them; this folder has several subdirectories of importance. In the ``\textit{Raw}'' subdirectory, the raw annotations outputted by PAGAN for each session can be found. The ``\textit{Raw Combined}'' subdirectory contains the raw data assembled into a single CSV for each task, and the ground truth signals for the QA test (see \textbf{Methods}). The ``\textit{Processed}'' subdirectory contains the processed annotation data for the QA tasks and the Engagement task for all sessions in NumPy file format. These processed files consist of a Python dictionary containing the following hierarchy: data is sorted into sessions, which in turn are sorted into participants, and further sorted by game where the participants' annotations can be found. The ``\textit{Latent Extraction}'' subdirectory contains Python scripts to extract the latents from the audiovisual data found in the ``\textit{Stimuli}'' directory. We also include examples of extracted latents for a 3-second time window that can be used for further analysis. Finally, the ``\textit{Analysis}'' directory contains the scripts used to conduct the analysis in this paper, and includes a ``\textit{Processed}'' subdirectory with annotation data with outliers removed (see \textbf{Methods}).

\section*{Technical Validation}

\textbf{Annotator Quality Assurance:}
We first illustrate the reliability of our annotators by computing their average QA scores across our visual and auditory QA tasks (see \textbf{Methods}). Since both QA tasks involve an annotation task where the ground truth (screen brightness and audio frequency respectively) is known in advance, we use the Signed Differential Agreement (SDA) metric \cite{narayanan2020green} to measure similarity. We chose SDA as it has been proven effective in our previous QA study \cite{barthet2023knowing} and, due to its bounded nature between -1 and 1, is an intuitive metric for selecting a filter threshold. Across all participants and both QA tasks, the average SDA score was $0.43\pm0.15$, meaning that annotators correctly annotated $71.5\%$ of all time windows among QA signals. If we used a QA filtering threshold value similar to previous work\cite{barthet2023knowing} (SDA = 0, or $50\%$ accuracy), only 3 annotators (P6 and P7 from Session 2, P17 from Session 4) would be removed due to poor QA performance. This verifies that almost all our annotators understood the annotation process and could accurately produce annotations before moving on to the engagement task. For the purposes of the technical validation of this paper, however, no participants are removed through this QA filtering in order to assess all raw data included in the dataset.

\begin{table}[t]
\centering
\footnotesize
\begin{tabular}{
c@{\extracolsep{3pt}}c@{\;}c@{\;}c@{\;}c@{\;}c@{\extracolsep{15pt}}
c@{\extracolsep{3pt}}c@{\;}c@{\;}c@{\;}c@{\;}c@{\extracolsep{15pt}}
c@{\extracolsep{3pt}}c@{\;}c@{\;}c@{\;}c@{\;}c@{\extracolsep{15pt}}
c@{\extracolsep{3pt}}c@{\;}c@{\;}c@{\;}c@{\;}c@{\;}c@{\;}c@{\;}c@{\;}}


\cmidrule[1.4pt]{1-6} 
\cmidrule[1.4pt]{7-12} 
\cmidrule[1.4pt]{13-18} 
\cmidrule[1.4pt]{19-24}

\multicolumn{6}{c}{\textbf{Session 1}} & \multicolumn{6}{c}{\textbf{Session 2}} & \multicolumn{6}{c}{\textbf{Session 3}} & \multicolumn{6}{c}{\textbf{Session 4}} \\ 
\cmidrule{1-6} \cmidrule{7-12} \cmidrule{13-18} \cmidrule{19-24}

& P1 & P2 & P3 & P4 & P5 & 
& P6 & P7 & P8 & P9 & P10 & 
& P11 & P12 & P13 & P14 & P15 & 
& P16 & P17 & P18 & P19 & P20 \\

\cmidrule{1-6} \cmidrule{7-12} \cmidrule{13-18} \cmidrule{19-24}
P1 &  & 4.1 & 8.9 & 8.5 & 4.3 & P6 & & 10.8 & 10.9 & 12.5 & 10.3 & P11 &  & 8.7 & 10.8 & 8.0 & 9.9 & P16 &  & 7.3 & 9.4 & 8.9 & 9.4 \\
P2 & 4.1 &  & 9.1 & 7.8 & 3.8 & P7 & 10.8 & & 6.4 & 8.2 & 10.3 & P12 & 8.7 &  & 8.1 & 8.8 & 8.8 & P17 & 7.3 &  & 12.2 & 12.8 & 11.3 \\
P3 & 8.9 & 9.1 &  & 5.1 & 11.2 & P8 & 10.9 & 6.4 &  & 9.8 & 10.2 & P13 & 10.8 & 8.1 &  & 11.7 & 11.4 & P18 & 9.4 & 12.2 &  & 10.5 & 8.6 \\
P4 & 8.5 & 7.8 & 5.1 &  & 9.2 & P9 & 12.5 & 8.2 & 9.8 &  & 9.6 & P14 & 8.0 & 8.8 & 11.7 &  & 9.6 & P19 & 8.9 & 12.8 & 10.5 &  & 6.3 \\
P5 & 4.3 & 3.8 & 11.2 & 9.2 &  & P10 & 10.3 & 10.3 & 10.2 & 9.6 &  & P15 & 9.9 & 8.8 & 11.4 & 9.6 &  & P20 & 9.4 & 11.3 & 8.6 & 6.3 & \\

\cmidrule[1.2pt]{1-6} 
\cmidrule[1.2pt]{7-12} 
\cmidrule[1.2pt]{13-18} 
\cmidrule[1.2pt]{19-24}

\textbf{Avg.} & \textbf{6.4} & \textbf{6.2} & \textbf{8.6} & \textbf{7.6} & \textbf{7.1} & \textbf{Avg.} & \textbf{11.1} & \textbf{8.9} & \textbf{9.3} & \textbf{10.0} & \textbf{10.1} & \textbf{Avg.} & \textbf{9.4} & \textbf{8.6} & \textbf{10.5} & \textbf{9.5} & \textbf{9.9} & \textbf{Avg.} & \textbf{8.8} & \textbf{10.9} & \textbf{10.2} & \textbf{9.6} & \textbf{8.9} \\

\cmidrule[1.4pt]{1-6} 
\cmidrule[1.4pt]{7-12} 
\cmidrule[1.4pt]{13-18} 
\cmidrule[1.4pt]{19-24}

\end{tabular}
\caption{\label{tab:dtw_distancematrix} DTW distance matrices between all participants in the same session, averaged across 30 video stimuli per session.}
\end{table}

\textbf{Inter-Annotator Agreement:} 
We use DTW (see \textbf{Methods}) to create distance matrices for each video in each session, in order to calculate the inter-annotator agreement between annotators' normalised engagement traces on the same stimuli. We average those DTW values between pairs of participants on a per-session basis, since the same participant with the same identifier (e.g. P1) annotated every video in the same session. We report the average DTW distance across all stimuli (30) per session in Table \ref{tab:dtw_distancematrix}, using different participant identifiers because different participants annotated different sessions. We observe that there are differences between annotators; only annotators in Session 1 were more in agreement. However, annotators are rarely consistently disagreeing with all other annotators; the most obvious instances of this are P6 (with DTW distances over 10 with all other annotators in Session 2) and P17 (with DTW distances over 11 with three other annotators in Session 4). We note that both annotators performed poorly in the QA tasks as well, and would have been removed via an SDA cut-off on the two QA tasks (see above). In general, there is a statistically significant ($p<0.05$) negative Pearson correlation between average DTW distance of each participant's engagement traces with all others in the same session and SDA score averaged from the two QA tasks ($r=-0.56$). This is a promising finding for the quality assurance offered by the additional QA tasks, as the closer the annotators are to the known ground truth in those two tasks (an SDA score near 1), the less likely they are to disagree with other annotators in the more challenging engagement annotation task.

\textbf{Outliers per participant:}
We detect outliers on the engagement traces (see \textbf{Methods}) based on the KDE of all DTW matrices of all videos in the same session, and report results with two filters for outlier removal. When using the $90\%$ filter (see \textbf{Methods}), 60 out of 600 annotations are removed (15 per session),  whilst with the $80\%$ filter 120 annotations are removed (30 per session). Interestingly, while P7 would be removed due to poor QA task performance, none of their 30 traces are removed as outliers for the 90\% filter and only 2 traces are removed as outliers for the 80\% filter. On the other hand, P17 has the most outliers of all 20 participants: 11 of their 30 traces are removed as outliers for the 90\% filter and an extraordinary 16 out of 30 removed for the 80\% filter. Other participants with many removed outliers are P4 (with 9 outliers for 80\% filter) and P3, P13, P20 (each with 8 outliers for 80\% filter). The average SDA score from the two QA tasks is not significantly correlated ($p>0.05$) with the number of outliers removed: $r=-0.33$ for removed outliers per participant with the 90\% filter and $r=-0.25$ for outliers with the 80\% filter.

\textbf{Outliers per game and video:}
It is worthwhile to observe which game stimuli were more prone to inter-annotator disagreements, which would lead to more outliers removed through the above filtering process. After applying the $90\%$ filter, outliers were removed from 26 out of the 30 games. The games with no outliers were \textit{Apex Legends}, \textit{Wolfram}, \textit{Medal of Honour} (PS1) and \textit{Superhot}. The games with the most outliers were \textit{Outlaws}, \textit{Counter-Strike 16}, and \textit{Wolfenstein 3D} with 4 outliers each (out of 20 annotation traces across 4 videos of the same game). After applying the $80\%$ filter,  outliers were removed from every game except \textit{Apex Legends}. The game with the most outliers when using the $80\%$ filter was \textit{Operation Body Count,} with 9 outliers removed (out of 20 traces). The consistency of \emph{Apex Legends} across sessions and annotators is likely due to intertwined factors, such as its highly stylised graphics, strong emphasis on audio cues, and quick well-defined changes in tempo (e.g. alternating between fast-paced team fights and slow-paced periods of healing, looting, exploration, etc.). Finally, observing individual videos with the most outliers after applying the $90\%$ filter, Session 1's \textit{Outlaws} and Session 4's \textit{Counter-Strike 16} had 3 outliers removed each (out of 5 traces each). When using the stricter $80\%$ filter, Session 4's \textit{Superhot} video had all of its annotations classified as outliers and removed, indicating that this video had particularly poor inter-rater agreement compared to others in that session. This aggressive filtering process could evidently prove problematic, as in the case of this video, no data remain for affect modelling. For this reason, the data records contain all raw and processed traces and Python scripts that would allow researchers to perform their own filtering process or apply different KDE thresholds depending on their goals.


\section*{Code availability}
The data have been deposited into an Open Science Framework repository \cite{game_vibe_osf} and is available for public use. The dataset can be managed, visualised and pre-processed using  Python files files.

\bibliography{bibliography}

\section*{Acknowledgements} 
This research is supported by the Foundation.AI (Foundation Modelling of Games via Artifcial Intelligence) research project funded by the Ministry for Education, Sport, Youth, Research and Innovation in Malta, as well as by the Malta Council for Science and Technology through the SINO-MALTA Fund 2022, Project OPtiMaL. Makantasis was supported by Project ERICA (GA: REP-2023-36) fnanced by the Xjenza Malta (Malta Council for Science \& Technology - MCST), for and on behalf of the Foundation for Science and Technology,through the FUSION: R\&I Research Excellence Programme.

\section*{Author contributions statement}
G.N.Y conceived the experiments, M.B, M.K and K.P conducted the experiments under the supervision of K.M, A.L and G.N.Y, and M.B and M.K performed the analysis of results. All authors discussed the results, and all authors contributed equally to the final manuscript.

\section*{Competing interests}
The authors declare no competing interests.

\end{document}